\documentclass[aps,superscriptaddress,showpacs,amsmath,amssymb,amsfonts,altaffillsymbol,twocolumn,notitlepage,pre]{revtex4-2}

\usepackage{graphicx}
\usepackage{float}
\usepackage{dcolumn}
\usepackage{bm}
\usepackage{hyperref}
\usepackage[table]{xcolor}
\usepackage{booktabs}

\newcommand{\Up}{\mathbf{U}}
\newcommand{\Dn}{\mathbf{D}}

\begin{document}
\title{Topology of the energy landscape of sheared amorphous solids and the irreversibility transition}

\date{\today}

\author{Ido Regev}
\email[Corresponding author: ]{regevid@bgu.ac.il}
\affiliation{Department of Solar Energy and Environmental Physics,
Jacob Blaustein Institutes for Desert Research, Ben-Gurion University of the Negev, Sede Boqer Campus 84990, Israel}
\author{Ido Attia}
\affiliation{Jacob Blaustein Institutes for Desert Research, Ben-Gurion University of the Negev, Sede Boqer Campus 84990, Israel}

\author{Karin Dahmen}
\affiliation{Department of Physics, University of Illinois at Urbana-Champaign, 1110 West Green Street, Urbana, IL 61801, USA}

\author{Srikanth Sastry}
\affiliation{Jawaharlal Nehru Centre for Advanced Scientific Research, Jakkar Campus, 560064 Bengaluru, India}

\author{Muhittin Mungan}
\email[Corresponding author: ]{mungan@iam.uni-bonn.de}
\affiliation{Institut f\"{u}r angewandte Mathematik, Universit\"{a}t Bonn, Endenicher Allee 60, 53115 Bonn, Germany}

\begin{abstract} 
Recent experiments and simulations of amorphous solids plastically deformed by an oscillatory drive have found a surprising behavior - for small strain amplitudes the dynamics can be reversible, which is contrary to the usual notion of plasticity as an irreversible form of deformation. This reversibility allows the system to reach limit-cycles in which plastic events repeat indefinitely under the oscillatory drive.  It was also found that reaching reversible limit-cycles, can take a large number of driving cycles and it was surmised that the plastic events encountered during the transient period are not encountered again and are thus irreversible.
Using a graph representation of the stable configurations of the system and the plastic events connecting them, we show that the notion of reversibility in these systems is more subtle. We find that reversible plastic events are abundant, and that a large portion of the plastic events encountered during the transient period are actually reversible, in the sense that they can be part of a reversible deformation path. More specifically, we observe that the transition graph can be decomposed into clusters of configurations that are connected by reversible transitions. These clusters are the strongly connected components of the transition graph and their sizes turn out to be power-law distributed. The largest of these are grouped in regions of reversibility, which in turn are confined by regions of irreversibility whose number proliferates at larger strains. 
Our results provide an explanation for the irreversibility transition - the divergence of the transient period at a critical forcing amplitude. The long transients result from transition between clusters of reversibility in a search for a cluster large enough to contain a limit-cycle of a specific amplitude. For large enough amplitudes, the search time becomes very large, since the sizes of the limit cycles become incompatible with the sizes of the regions of reversibility. 
\end{abstract}

\maketitle

\section{Introduction and Summary}

Understanding the response of a configuration of interacting particles in a  disordered solid to an externally imposed force is one of the main challenges currently facing researchers in the fields of soft matter physics and rheology \cite{bonn2017yield,keim2019memory}.  As an amorphous solid adapts to the imposed forcing, it starts to explore its complex energy landscape which gives rise to rich dynamics \cite{szulc2020forced,schinasi2020annealing}. 
One example of such dynamics is the response to an oscillatory driving, which for small amplitudes, can lead to cyclic response: a repeated sequence of configurations whose period is commensurate with that of the driving force \cite{regev2013onset,fiocco2013oscillatory,keim2014mechanical,schreck2013particle,priezjev2016reversible,royer2015precisely,bandi2018training}.  Such cyclic responses encode information and possess ``memory''  about the forcing that caused them. Memory effects of this kind have been observed experimentally \cite{keim2020global}, as well as numerically \cite{fiocco2014encoding,regev2013onset}, in cyclically driven (sheared) amorphous solids, colloidal suspensions \cite{keim2019memory}, and other condensed matter systems, such as superconducting vortices
and plastically deformed crystals \cite{brown2019reversible,sethna1993hysteresis,pine2005chaos,corte2008random,Keimetal2011,ni2019yield}. 
Cyclic response is important in many applications of plastic deformation such as fatigue experiments and the stability of geophysical structures. Large Amplitude Oscillatory Shear (LAOS) is used extensively to characterize the rheological properties of soft materials\cite{HYUN20111697}.

An important feature of cyclic response in amorphous solids is that for small  shear amplitudes, the steady state response includes plastic events that keep reoccurring in consecutive driving cycles and are in this sense reversible. 
However, before the system settles in a cyclic response, it typically undergoes a transient period in which the dynamics 
is not repetitive and the plastic events can thus be regarded as irreversible. As the amplitude of shear is increased, transients become increasingly longer and eventually, at a critical strain amplitude, the dynamics becomes completely irreversible. Here we will consider this critical strain to be the yield strain \cite{regev2015reversibility,fiocco2013oscillatory} though it is
sometimes referred to as the irreversibility transition.

Since both reversible and irreversible plastic events involve 
particle rearrangements, it is not clear what distinguishes one from the other \cite{regev2015reversibility,leishangthem2017yielding,priezjev2018yielding,mangan2008reversible,das2020unified,morse2020differences,kawasaki2016macroscopic,regev2018critical,nagamanasa2014experimental,mobius2014ir}. Recently, we have shown that in the athermal, quasi-static (AQS) regime we can rigorously describe the dynamics of driven disordered systems in terms of a directed state-transition graph \cite{munganterzi2018, munganwitten2018, mungan2019networks}. The nodes of this graph, the mesostates, correspond to collections of locally stable particle configurations that transform into each other under applied shear via purely elastic deformations. The edges of the graph therefore describe the plastic events.  Furthermore, we have demonstrated that such transition graphs can be readily extracted from molecular simulations of sheared amorphous solids \cite{mungan2019networks}. The ability to link the topology of the AQS transition graph with dynamics has provided a novel means of probing the complex energy landscape of these systems. Here we show that analysis of such graphs in terms of their strongly connected components (SCCs) \cite{barrat2008dynamical} allows us to distinguish between reversible and irreversible events and better understand the organization of memory in these materials. 

In the AQS networks of sheared amorphous solids, SCCs correspond to sets of mesostates connected by plastic deformation pathways such that each mesostate in the SCC is reachable from each other mesostate in the SCC by a plastic deformation path. Due to this property,
a plastic transition which is part of these paths can be reached arbitrarily many times and is reversible.
Reversible plastic events are thus events that connect states within an SCC. Conversely, irreversible plastic events are  transitions between states in different SCCs, since by their definition, transitions between mesostates belonging to different SCCs cannot be reversed. The ability to identify reversible plastic events as events inside SCCs and irreversible plastic events as transitions between SCCs allows us to better understand the transient and reversible dynamics of amorphous solids. At the same time, this distinction facilitates  comparing the properties of the corresponding plastic events.  
We observe that changes in energy and stress during irreversible events are significantly larger than for reversible events. While many irreversible transitions occur at high stresses and energies associated with yielding, we also find a significant amount of irreversible transitions occurring at much lower stresses and energies.

We further study the properties of SCCs and find that their overall size distribution follows a power-law. 
For strains near and above yield, very small SCCs proliferate. Since the plastic events associated with cyclic response are reversible, they must be confined to a single SCC. The statistics of SCC sizes thus provides an estimate of the memory retention capability and its dependence on the strain amplitude of the driving. Furthermore, these findings also shed light on the mechanisms giving rise to 
the long transient dynamics observed in cyclically sheared amorphous solids. 
We show that reversible plastic events are dominant up to a strain of about $\gamma_{\rm rev} = 0.085$, which is below the yield strain  $\gamma_y = 0.135$ in this system.  For strains above $\gamma_{\rm rev}$ and approaching yielding,  irreversible plastic events become increasingly dominant. This finding 
 suggests that there is a change in the dynamic response of these systems as the driving crosses from the below yield to the near yield regime around $\gamma_{\rm rev} = 0.085$. 
Indeed, we find that in the sub-yield regime $\gamma < \gamma_{\rm rev}$, large SCCs are readily available and the transient to a limit-cycle is largely constrained by finding the right one, {\em i.e.} a response where all plastic transitions are reversible and thus confined to the same SCC. On the other hand at the near-yield regime, $\gamma_{\rm rev} < \gamma < \gamma_y$, the SCC size does matter. This regime is characterized by small SCCs and hence SCCs of the required size are rare. As a result, the transient dynamics is dominated by a search for an SCC of the appropriate size.   
\begin{figure*}[t]
\begin{center}
\includegraphics[width=2\columnwidth]{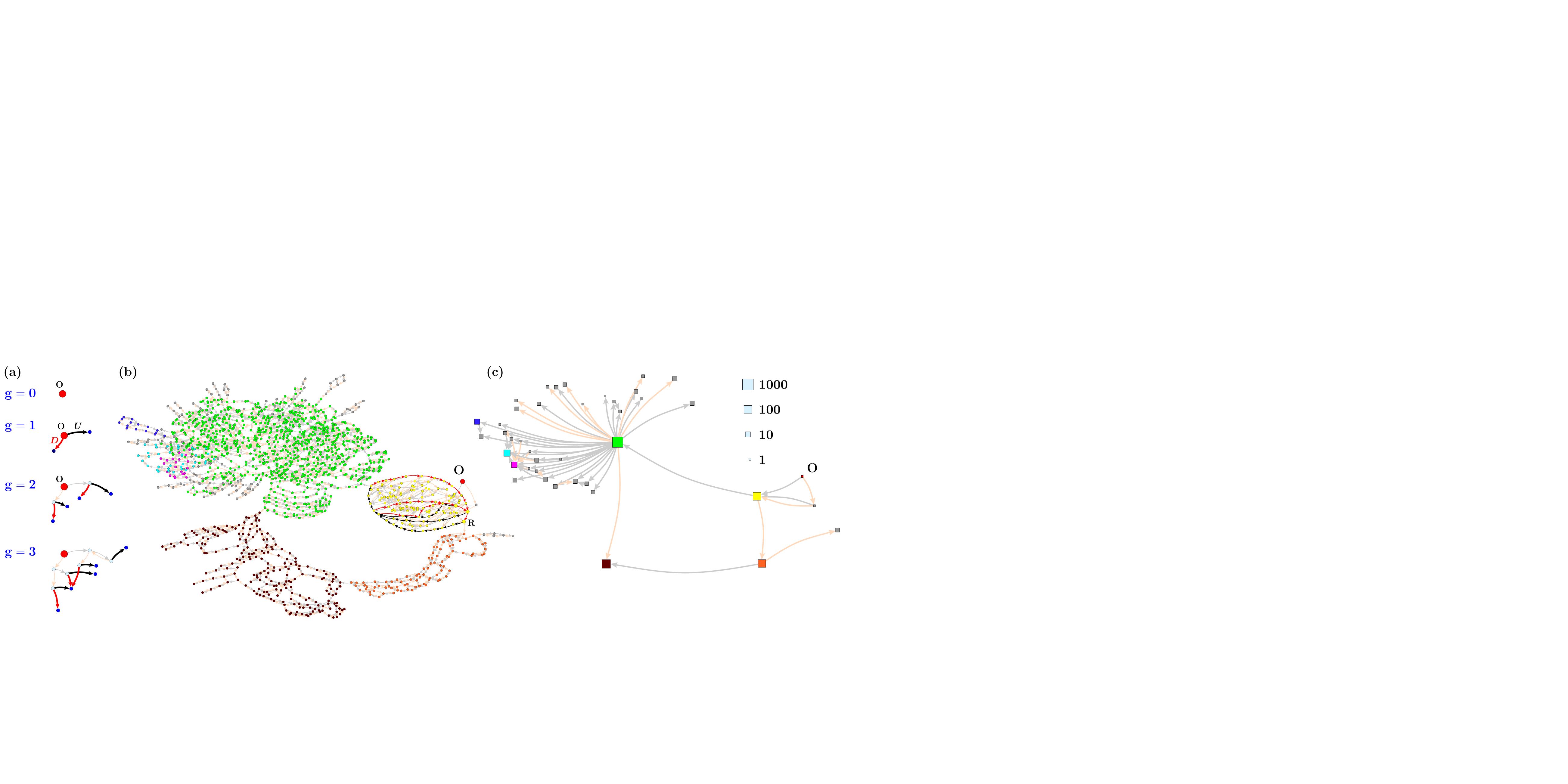}
\caption{(a) Illustration of the construction of a catalogue of mesostates starting from the reference state $O$ at generation $g = 0$. Transitions in black/gray (red/orange) designate $\Up$- ($\Dn$-) transitions. Under $\Up$- and $\Dn$-transitions we obtain $2, 4$, and $5$ new meostates at generations $ g= 1,2$, and $3$, respectively. Transitions leading to new mesostates at each generation have been highlighted. (b) A mesostate transition graph generated from an initial configuration $O$ (marked as a red dot) with several strongly connected components (SCCs) highlighted in different colors. The largest 6 SCCs have sizes, 929 (green), 222 (brown), 115 (yellow), 90 (orange), 37 (cyan), and 20 (purple). Transitions within an SCC correspond to reversible plastic events, since for any deformation path connecting two states in an SCC, by definition there is also a reverse path. Irreversible plastic events are transitions between states belonging to different SCCs.   
(c) The inter-SCC graph is a compressed representation of the graph in (b), showing the SCCs as squares with colors that correspond to the colors in (b). The arrows connecting the SCCs are the irreversible plastic events and the inter-SCC graph is therefore acyclic.  }
\label{Fig1}
\end{center}
\end{figure*}
\section{Results}

\subsection{Mesostates, AQS state transition graphs, and mutual reachability}

Consider the athermal dynamics of an amorphous solid being subject to shear strain along a fixed direction. After its initial preparation, before the system is subject to any external forcing, it is in a local minimum of its potential energy. As we increase the strain in a slow and adiabatic manner, the energy landscape deforms and the position of the local energy minimum in configuration space changes. For a range of strains that is dependent on the particle configuration, the amorphous solid adapts by purely elastic deformation to the strain increments. This elastic response lasts until we reach a value of the strain where the particle configuration attained ceases to be a local energy minimum and thus becomes unstable. Increasing the strain further, the system relaxes into a new local energy minimum and this constitutes a plastic event. Thus, given a locally stable configuration of particles, there exists a range of strains, applied in the positive and negative directions, over which an amorphous solid adapts to changes in the applied strain in a purely elastic manner and which is punctuated on either end by plastic events. In \cite{mungan2019networks} we have called such a contiguous collection of locally stable equilibria a ``mesostate''. Thus with each mesostate $A$ we can associate a range of strain values $(\gamma^-[A], \gamma^+[A])$, over which the locally stable configurations transform elastically into each other and that is limited by plastic events at $\gamma^\pm[A]$. When a plastic event occurs, the system reaches a new, locally stable, configuration which must necessarily belong to some other mesostate $B$. Since mesostate transitions are triggered at either end of the stability interval $(\gamma^-[A], \gamma^+[A])$, we call transitions under strain increase and decrease $\Up$-, respectively, $\Dn$-transitions. For example,  if mesostate $B$ is reached under a $\Up$-transition from $A$, we write this symbolically  as $B = \Up A$. The mesostate transitions under strain increases and decreases have a natural representation as a directed graph, the AQS state transition graph. Here each vertex is a mesostate and from each vertex we have two outgoing directed transitions, namely one under $\Up$ and the other under $\Dn$. As explained above, in the context of sheared amorphous solids, the transitions of the AQS graphs correspond to purely plastic events. These events can be traced back to localized regions in the sample, the soft-spots, where a small number particles undergo a rearrangement. 
In the simplest picture, soft spots can be thought of as two level hysteretic elements
\cite{manning2011softspot,falk1998dynamics}, which interact with each other via Eshelby-type long range elastic deformation fields \cite{maloney2006amorphous}. 

Since the AQS transition graph represents the plastic deformation paths under every possible history of applied shear along a fixed axis, the dynamic response of the amorphous solid will be encoded in the graph topology. The connection with soft-spot interactions was already explored in \cite{mungan2019networks}, and our aim here is to explore the implications of graph topology on the dynamics. To this end, we perform a decomposition of the graph into its strongly connected components. This decomposition is based on the relation of mutual reachability of mesostates, which is defined as follows: two mesostates $A$ and $B$ are said to be mutually reachable if there is a sequence of $\Up$ and $\Dn$ transitions that lead from $A$ to $B$ and back from $B$ to $A$. Mutual reachability is an equivalence relation: if $A$ and $B$ are mutually reachable and $B$ and $C$ are mutually reachable, then $A$ and $C$ are also mutually reachable. Thus mutual reachability partitions the set of mesostates of the AQS transition graph into (disjoint) sets of mutually reachable states. In network theory such sets are called strongly connected components (SCCs) \cite{barrat2008dynamical}. 

\subsection{AQS transition graphs from simulations}
As we have shown in \cite{mungan2019networks}, it is possible to extract AQS state transition graphs from simulations of a sheared amorphous solid. 
The details of the construction of such graphs can be found in Appendix \ref{supp:sample_prep} and \ref{supp:catalog_extraction}, as well as the Supporting Material of ref. \cite{mungan2019networks}.  Here we summarize the main procedure and our data. We start with an initial stable particle configuration belonging to a mesostate $O$, which we call the reference state, and we assign $O$ to generation $g = 0$. We then determine its range of stability $\gamma^\pm[O]$, as well as the mesostates $\Up O$ and $\Dn O$ that it transits into. The latter are the mesostates of generation $g = 1$. Proceeding generation by generation, and identifying mesostates that have been encountered at a previous generation we can assemble a catalog of mesostates, which (i) lists the stability range of each mesostate, and (ii),  identifies the mesostates that these transit into under $\Up$- and $\Dn$-transitions.  Fig.~\ref{Fig1}(a) illustrates the initial stages of the catalog acquisition.  
We have extracted from numerical simulations $8$ catalogs, each corresponding to a different initial configuration quenched from a liquid. These catalogs contain a total of nearly $400$k mesostates  and we identified the SCCs that they belong to. Table \ref{tab:g002data} in the Appendix \ref{supp:catalog_extraction} summarizes our data.  Fig.~\ref{Fig1}(b) shows a portion of an AQS state transition graph obtained from catalog $\#1$ of the data set. The excerpt shown contains $1542$ mesostates. The reference state $O$, containing the initial configuration, is marked by a big circle (in red) and nodes belonging to the same SCC have the same color. SCCs with less than 15 nodes are shown in dark gray. The largest 6 SCCs shown have sizes, 929 (green), 222 (brown), 115 (yellow), 90 (orange), 37 (cyan), and 20 (purple). 
\begin{figure}[t]
\begin{center}
\includegraphics[width=\columnwidth]{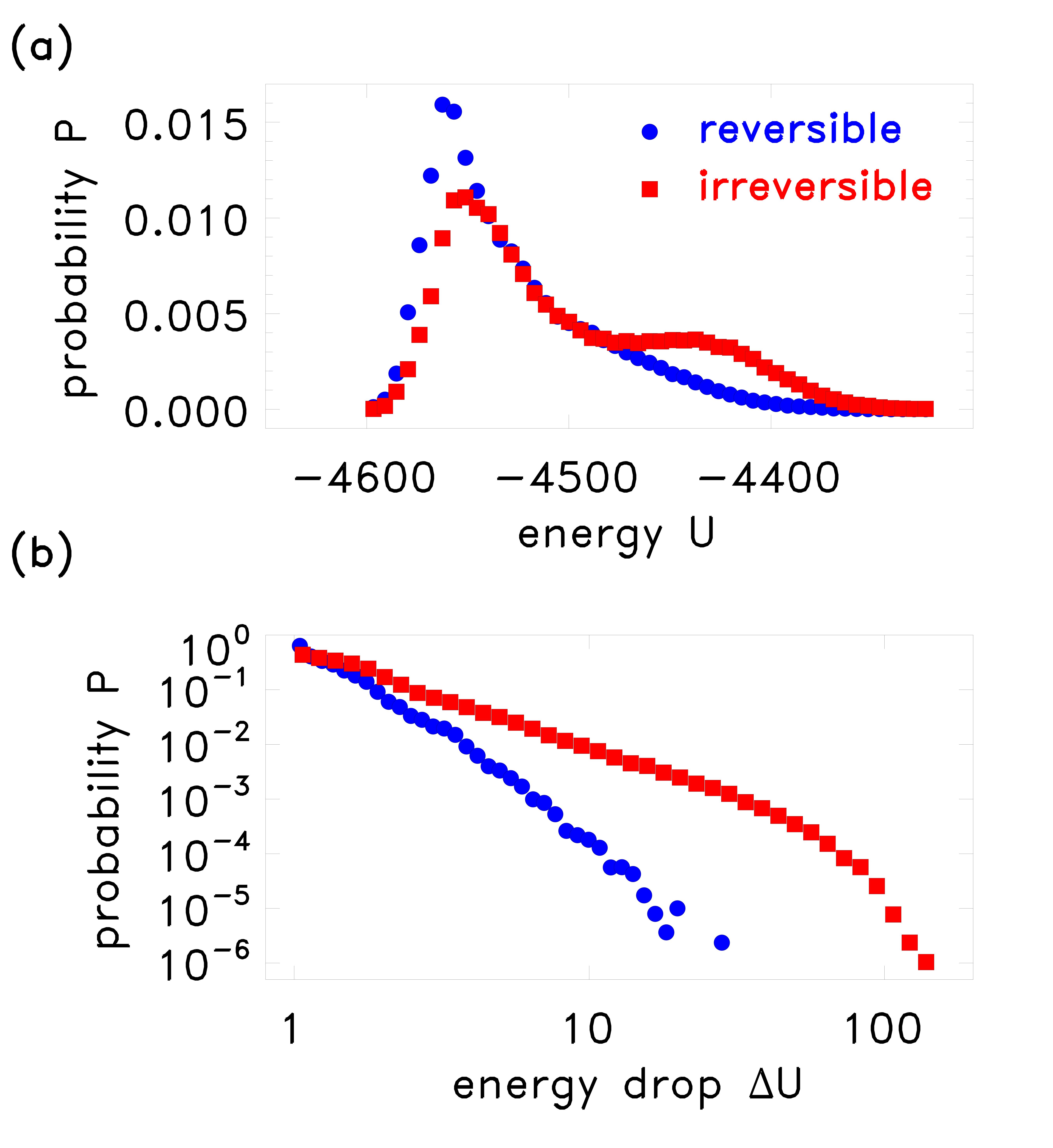}
\caption{
(a) Normalized distributions of energies at the onset of reversible (blue dots) and irreversible (red squares) plastic events (transitions). 
(b)  Normalized distributions of the energy drops during reversible (blue dots) and irreversible (red squares) plastic events (transitions). 
The results in this figure combine data sampled from all $8$ catalogs.
}
\label{Fig2a}
\end{center}
\end{figure}
\begin{figure*}[t]
\begin{center}
\includegraphics[width=2\columnwidth]{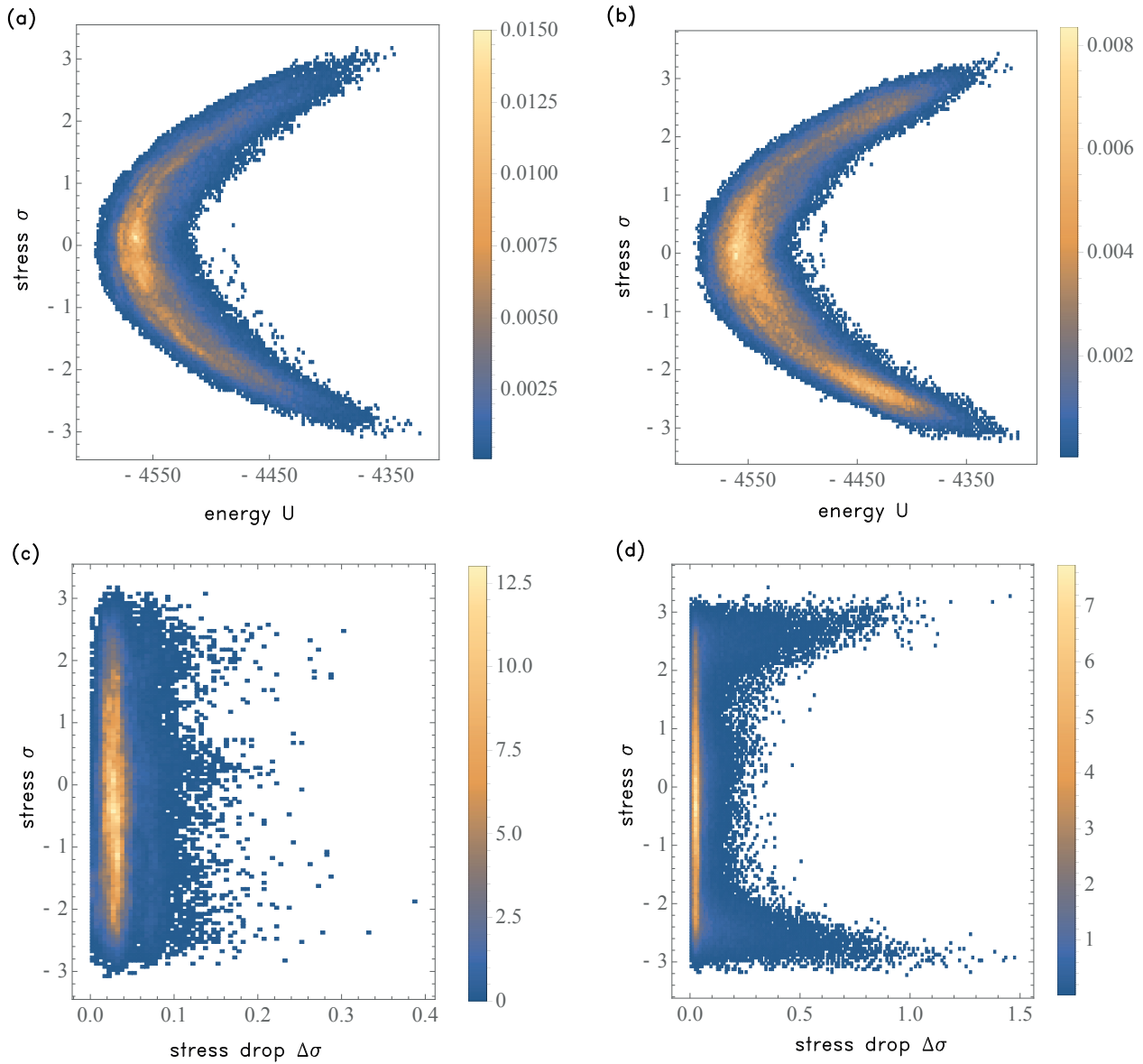}
\caption{
(a,b) Density plot of stress vs. the energy at the onset of reversible (a) and irreversible (b) plastic events. The overall parabolic shape of the scattered points, corresponds to the bulk elastic response of the samples.  
(c,d) The stress drop after a plastic event vs. the stress at the onset of the plastic event for reversible (c) and irreversible (d) transitions. The color bars to the right depict the color-coding of the density from low (dark/blue) to high (bright/yellow). 
The results in this figure combine data sampled from all $8$ catalogs}.
\label{Fig2b}
\end{center}
\end{figure*}
\subsection{Reversible and Irreversible Plastic Events}
The partition of the mesostates of an AQS transition graph into SCCs allows us to identify two types of transitions: transitions within the same SCC and transitions connecting different SCCs. The former transitions are plastic deformations that can be reverted, since mutual reachability assures that for any transition from $A$ to $B$ there exists a sequence of transitions from $B$ to $A$. We will therefore call these transitions {\em reversible} \footnote{Note that this is not reversibility in the thermodynamic sense, since plastic events involve energy dissipation.}. On the other hand, transitions between two different SCCs must necessarily be {\em irreversible}: there is a plastic deformation path from a mesostate in one SCC to a mesostate in another SCC, but there is no deformation path back. If there had been one, these two states would have been mutually reachable, and therefore part of the same SCC. Further details on identifiying transitions as reversible are given in Section \ref{supp:scc_identification} of the the Appendix. 
We can condense the transition graph by collapsing all states belonging to an SCC into a single vertex so that only transitions between SCCs remain \cite{corominas2013origins}, {\em i.e.} the irreversible transitions. The graph obtained in this way is the inter-SCC graph, and by construction, this graph is acyclic, {\em i.e.} it cannot contain any paths that lead out of and return to the same SCC. Fig.~\ref{Fig1}(c) shows the inter-SCC graph obtained from the mesostate transition network shown in panel (b). The size of the vertices represents the size of the respective SCCs with a logarithmic scaling as indicated in the legend to the right of the figure. The color and placement of the SCC vertices follows those of panel (b). 

Since the SCC decomposition allows us to distinguish reversible from irreversible plastic events, we can use it to compare their properties. 
In Figs~\ref{Fig2a} and \ref{Fig2b} we compare the statistics of reversible and irreversible events across the entire $8$ catalogues. 
In Fig~\ref{Fig2a}(a) we show the energies at the onset of reversible (blue dots) and irreversible (red squares) plastic events.
We see that while reversible events occur predominately at low energies, the distribution for irreversible events is bimodal: there is a concentration of events at low energies and another concentration at high energies. Fig~\ref{Fig2b}(a,b) shows density plots of the stress at which reversible and irreversible plastic events occurs as a function of the energy. We can see that the secondary peak of irreversible transitions at higher energies correspond to stresses $\sigma$ close to and above the yield stress (the stress at the yielding/irreversibility transition), which is $\sigma_y \sim 2.5$ in units of the simulation and that reversible events are much scarcer in this region. In Fig~\ref{Fig2a}(b)
we compare the energy drops due to reversible and irreversible plastic events. We can see that both exhibit power-law behavior.  The irreversible events, while showing a strong cutoff, give rise to much larger energy drops in general and correspond to large collective particle rearrangements (avalanches). In Fig~\ref{Fig2b}(c,d) we show a density plot of the stress drops $\Delta\sigma$ and stresses $\sigma$ associated with reversible and irreversible plastic events, respectively. The figure reveals that the events accompanied by large stress avalanches are concentrated close to and above the yield stress and exhibit a secondary peak in the density plot of the irreversible events.
While it is obvious that close to yielding the system experiences a large number of large irreversible events, the figures also clearly shows the presence of a large number of irreversible events with small stress drops at stresses much below yield. In the following we shall argue that these events play a role in the transient dynamics observed in simulations under oscillatory shear at sub-yield strain amplitudes \cite{fiocco2013oscillatory,regev2013onset,kawasaki2016macroscopic,regev2018critical}. 
\begin{figure*}[t!]
\begin{center}
\includegraphics[width=2\columnwidth]{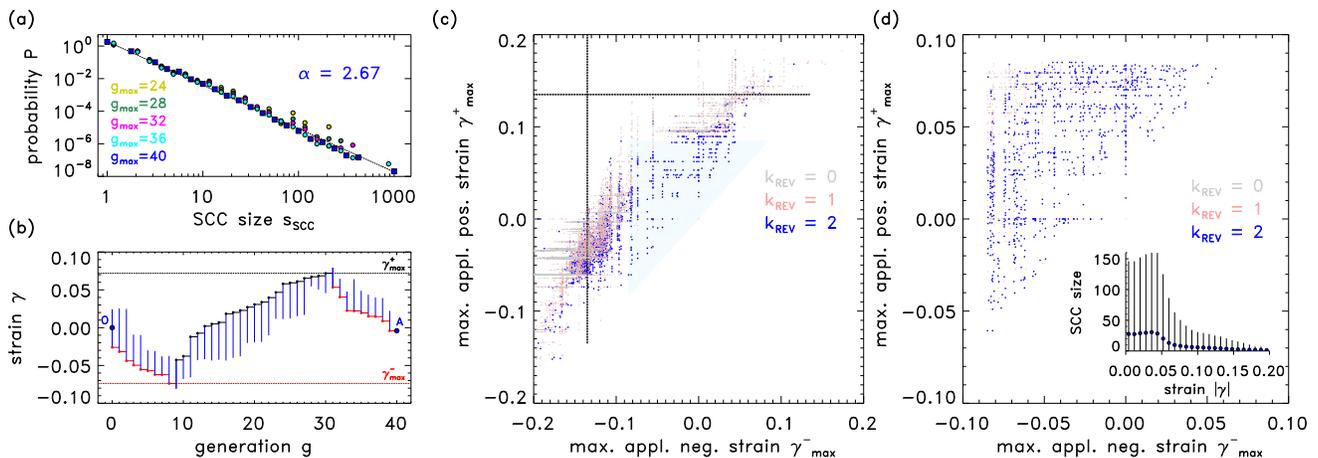}
\caption{ 
(a) The  SCC size distribution taken from the full $8$ catalogues (in blue) exhibits a heavy tail. The solid line is a power-law exponent $2.67$ and serves as a guide to the eye. 
Colors other than blue correspond to distributions derived from the same catalogues but only up to a maximal generation number of $24,28,32,36$ demonstrating that the distribution of SCC sizes becomes stable for networks significantly smaller than the ones used to calculate the exponent.
(b) Plastic deformation history leading from the initial state $O$ to a mesostate $A$ of the catalog after $g = 40$ plastic events. 
Each vertical blue line is an intermediate mesostate $P$ with its stability range $(\gamma^-[P],\gamma^+[P])$, while the horizontal 
 line segments in black ($\Up$) and red ($\Dn$) that connect adjacent mesostates indicate the strains at which the corresponding plastic events occurred. 
 For each mesostate $A$ and deformation history, we can identify the largest and smallest strains under which a $\Up$-, respectively $\Dn$-transition occurred, $\gamma^\pm_{\rm max}$, as illustrated by the extended horizontal lines. 
(c) Deformation path history dependence of $k_{\rm REV}$: each dot represents a mesostate of catalog $\#1$. The coordinates of each dot represent the largest positive and negative strains $\gamma_{\rm max}^\pm$, {\em cf.} panel (b), that were required to reach a specific mesostate, while  their color represents how many reversible transitions  $k_{\rm REV}=0, 1$, or $2$, go out of it, as indicated in the legend. The location of the yield strain in both positive and negative direction have been marked by dotted vertical and horizontal lines. The region highlighted by the light blue triangle contains the set of all mesostates that can be reached without ever applying a shear strain whose magnitude exceeds $\vert \gamma_{\rm max}^\pm \vert = 0.085$. 
The prevalence of mesostates with $k_{\rm REV} = 2$ (blue dots) inside this region, implies that mesostates reached by applying strains whose magnitudes remain below $0.085$ undergo predominantly reversible transitions, {\em i.e.} lead to mesostates that are part of the same SCC.
(d) Scatter plot of the mesostates with $\vert \gamma_{\rm max}^\pm \vert \le 0.085$ across the $5$ catalogs with $40$ or more generation. As was the case for the single data set shown in panel (c) of this figure, the region $\vert \gamma_{\rm max}^\pm \vert \le 0.085$ shows a high degree of reversibility across all $5$ catalogs: the region contains $9298$ mesostates out of which $7728$ have $k_{\rm REV} = 2$ and $1194$ have $k_{\rm REV} = 1$ outgoing irreversible transitions.
Inset: mean SCC size that a mesostate belongs to, given that it is stable at some strain $\gamma$ calculated from all $8$ catalogs. Error bars represent the standard deviation of fluctuations around the mean. The figure shows that mesostates stable at large strains tend to belong to small SCCs. 
}
\label{Fig3}
\end{center}
\end{figure*}
\subsection{AQS transition graph topology}
Fig~\ref{Fig3}(a) shows the size distribution of SSCs extracted from all eight catalogs. The solid line is a power-law behavior with  exponent of $2.67$ and serves as a guide to the eye. We estimated the power-law exponent and its uncertainty using the maximum-likelihood method described in \cite{clauset2009power}, and by considering only the $24488$ SCCs with  sizes  $s_{\rm SCC} \ge s_{\rm min} = 4$. This choice was motivated by the empirical observation that small SCCs containing mesostates at the largest generations of  the catalog limit are more likely to increase in size, if the catalog is augmented by going to higher number of generations. The exponent depends on the choice of cutoff $s_{\rm min}$: for $s_{\rm min} = 1, 2, 3$, and  $4$, we obtain (number of data points indicated in parentheses) the exponents $2.033 \pm 0.003$ $(169049)$, $2.529 \pm 0.005$ $(81528)$, $2.60 \pm 0.01$ $(40021)$, and $2.67 \pm 0.01$ $(24488)$, respectively. The exponents for $s_{\rm min} = 2, 3$, and  $4$ fall all into a an interval between $2.5$ and $2.7$, while the exponent of $2.203$ obtained with the cut-off $s_{\rm min} = 1$  seems to be significantly different. In fact, as we will show shortly, close to yielding there is a proliferation of SCCs with size one and this affects the estimate of the exponent. Thus the distribution of SCC sizes is broad, following a power-law $s_{\rm SCC}^{-\alpha}$, with an exponent of about $\alpha = 2.67$ and with the main source of uncertainty in $\alpha$ coming from the choice of the lower cut-off $s_{\rm min}$. Fig~\ref{Fig3}(a) also compares this distribution against the distributions obtained by limiting the generation number in the catalogues to a maximal generation number. It is clear that the distribution does not change significantly.

Next, we ask for the  ``location'' of SCCs in the transition graph by looking for correlations between the plastic deformation history of a mesostate $A$ and the number  of reversible transitions 
that are going out of it, $k_{\rm REV}[A]$. Recall that each mesostate in our catalog is reached from the reference configuration $O$ by a sequence of $\Up$- and $\Dn$-transitions. We call this the plastic deformation history path of $A$, as illustrated in Fig.~\ref{Fig3}(b). Additional details on deformation history are provided in Section \ref{supp:scc_identification} of the Appendix.   
For each mesostate and deformation history path, we can identify the largest and smallest strains under which a $\Up$-, respectively $\Dn$-transition occurred, $\gamma^\pm_{\rm max}$. These values are indicated in Fig~\ref{Fig3}(b) by the horizontal dashed lines.
Fig~\ref{Fig3}(c) shows a scatter plot obtained from catalog $\#1$ of our data set. Here each dot corresponds to a mesostate $A$ that is placed at $(\gamma^-_{\rm max}[A],\gamma^+_{\rm max}[A])$. Since 
$\gamma^-_{\rm max}[A] < \gamma^+_{\rm max}[A]$, the dots are scattered above the central diagonal of the figure. The location of the yield strain $\gamma_y = 0.135$ of the sample is indicated by the dashed vertical 
and horizontal lines. We have color-coded the mesostates according to the number $k_{\rm REV}[A]$ of outgoing reversible transitions, with blue, light red and gray corresponding to $2$, $1$, and $0$ possible 
reversible transitions, respectively. Note that multiple mesostates can have the same values of the extremal strains $\gamma^\pm_{\rm max}$  
and hence will be placed at the same location in the scatter plot. In order to reveal correlations between the straining history and $k_{\rm REV}$, we have first plotted the data points for which $k_{\rm REV} = 2$, next those 
for which $k_{\rm REV} = 1$, and finally,  $k_{\rm REV} = 0$. In spite of this over-plotting sequence, there appears a prominent central ``blue'' region that is bounded by $\gamma^-_{\rm max} \ge -0.085$ and 
$\gamma^+_{\rm max} \le 0.085$. 
This region contains $1783$ mesostates out of which $1448$ have $k_{\rm REV} = 2$, $257$ have $k_{\rm REV} = 1$, while $78$ mesostates have $k_{\rm REV} = 0$. Thus  $88\%$ of the transitions out of these mesostates are reversible \footnote{Since the total number of states in this region is $1783$, the total number of outgoing transitions, being twice this number, is $3566$. The total number reversible transitions id $2\times1448 + 257 = 3153$. Hence the fraction of reversible transitions is $0.88$.}.
{\it States with a deformation history in which the magnitude of the applied strain never exceeded $0.085$ are therefore highly likely to deform reversibly} under $\Up$- or $\Dn$-transitions. 
Since irreversible transitions are rare in this region, 
and it is only these transitions that connect different SCCs, a further implication of this finding is that the mesostates in this regime must be organized in a small number of SCCs, and we therefore expect these to be large. 
Upon inspection, we find that the mesostates in this region belong to $199$ SCCs with the largest $8$ SCCs having sizes $s_{\rm SCC} = 929, 306, 222, 115, 90, 81, 33$, and  $30$ 
\footnote{The total size of the $199$ SCCs containing the $1783$ mesostates in this region is $2587$. Thus an additional $804$ mesostates belong to these SCCs, but are outside the reversibility region. All $2587$ mesostates turn out to be contained in the region  
$\gamma^-_{\rm max} \ge -0.095$ and $\gamma^+_{\rm max} \le 0.096$.}. The excerpt of the transition graph shown in Figs.~\ref{Fig1}(b) and \ref{Fig1}(c) contains some of these SCCs. We have verified that such reversibility regions are present in each of the 8 catalogs we extracted and with similar extents in strain $\vert \gamma^\pm_{\rm max} \vert \lesssim 0.085$. Fig.~\ref{Fig3}(d) shows a scatter plot of the mesostates 
with $\vert \gamma^\pm_{\rm max} \vert \le 0.085$ sampled from the $5$ catalogs with $40$ or more generations. This region contains $9298$ mesostates out of which $7728$ have only reversible outgoing transitions ($k_{\rm REV} = 2$), while 
for $1194$ mesostates one of the two transitions is reversible
($k_{\rm REV} = 1$). 

One prominent feature of the transition graph excerpt shown in Figs.~\ref{Fig1}(b) and (c) is the large hub-like SCC (green) with $s_{\rm SCC} = 929$ mesostates and an out-degree of $39$, {\em i.e.} $39$ $\Up$- or $\Dn$-transitions  to neighbouring SCCs. Hubs are a common feature of scale-free networks,  which typically emerge via a stochastic growth process of self-organization \cite{barabasi1999emergence,barrat2008dynamical}. Such networks are characterized by heavy-tailed degree distributions. 
Note that a mesostate transition graph is generated from a single disordered initial configuration of particles, by a deterministic process for the acquisition of mesostates and the identification of transitions between them. The initial configuration itself has been obtained from a liquid state through a quench to zero temperature. 
The transition graphs can therefore be regarded as quenched disordered graphs, linked via the catalog acquisition process to an ensemble of initial configurations extracted from the liquid state \cite{munganwitten2018,munganterzi2018}.
We have computed degree distributions of the inter-SCC graphs over the full catalog as well as when restricted to the reversibility regions. For example, among the $199$ SCCs associated with the reversibility region of catalog \#1,  Fig. \ref{Fig3}(c), the largest $8$ SCCs also have the largest degrees, given by $k = 39, 12, 7, 4, 3, 3, 2$, and $2$.  The reversibility regions of all $8$ catalogs display similar network features: in each of these we observe a few SCCs with large degrees that are superposed on a  background of a very large number of SCCs with very small degrees. Note that every SCC has to have at least two outgoing irreversible transitions, as explained in Section \ref{supp:scc_identification} of the Appendix. 
While these findings {\em per se}  do not rule out the possibility of a scale-free inter-SCC graph, here are not enough SCCs with large degrees in our catalogs to deduce a heavy-tailed degree distribution.

\begin{figure*}[t!]
\begin{center}
\includegraphics[width=2\columnwidth]{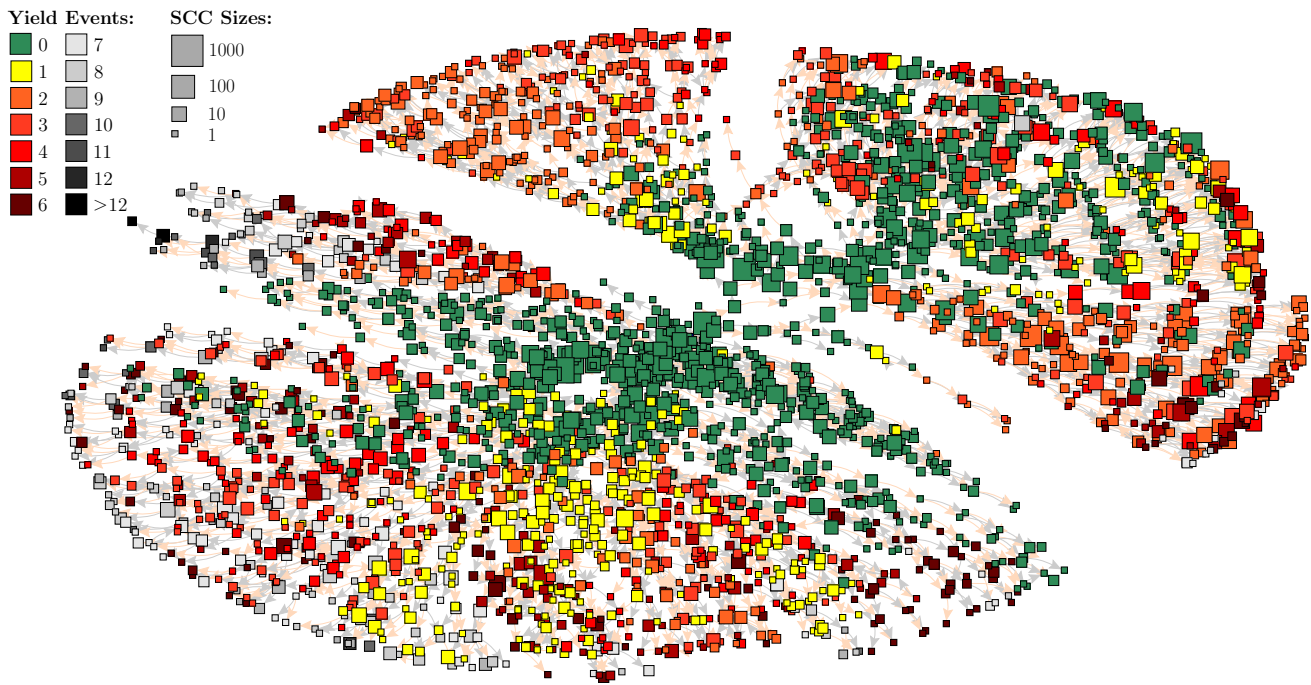}
\caption{ Large excerpt of the inter-SCC graph, {\em cf.} Fig.\ref{Fig1}(c), obtained from catalog \#1. Shown is the sub-graph of $3228$ SCCs (squares) that can be reached from the SCC of the initial state by at most $15$ inter-SCC transitions. These SCCs contain a total of $12790$ mesostates. The size of the SCCs is indicated by the legend in the top left corner of the figure. The coloring scheme of the SCCs indicates the (average) number of yield events in the plastic deformation history of the mesostates constituting the SCC, {\em i.e.} the number of mesostate transitions in their deformation history that occur at stresses of magnitude $2.5$ and higher. The figure shows patches of SCCs with the same number of yield events. Among these the 'green' patch of SCCs whose constituting mesostates have suffered no yield experience is dominant. Note that even for $2$ (orange) or more yield events there are relatively extended patches of SCCs of large sizes, such as the orange patch in the top left part of the graph. These findings suggest that the transition graph contains multiple reversibility regions, such as the one shown in Fig.~\ref{Fig3}(c), that differ only by the common history of past yield events of their constituent mesostates.
}
\label{Fig_interSCC_new}
\end{center}
\end{figure*}

The inset of Fig~\ref{Fig3}(d) shows the (conditional) average of the sizes of SCCs to which a mesostate $A$ belongs to, given that it is stable at some strain of magnitude  $\vert \gamma \vert $, 
{\em i.e.} we average over the sizes of SCCs  which a mesostate $A$ belongs to, and for which either $\gamma^-[A] < \vert \gamma \vert  < \gamma^+[A]$, or $\gamma^-[A] < -\vert \gamma \vert  < \gamma^+[A]$ holds. Further details are provided in Section \ref{supp:scc_identification} of the Appendix.
The vertical error bars show the standard deviations around the averages. For $ \vert \gamma \vert \lesssim 0.05$, the mean SCC size is around $30$.   
The distribution of SCC sizes in this region is very broad, as can be seen from the standard deviations, which are much larger than the mean values. 
For larger strains, the mean SCC size gradually drops to $1.2$, accompanied by increasingly smaller standard deviations. 
Since all states of a given catalog are reached from the same ancestral mesostate $O$ at zero strain, a mesostate whose strain history 
has never experienced a strain of magnitude larger than $ \gamma_{\rm max} $ must be stable at some strain $\gamma$ with $-\gamma_{\rm max} \le \gamma \le \gamma_{\rm max}$. Thus mesostates inside the reversibility region are stable at strain values that are also within these ranges. We therefore conclude that the mesostates in the reversibility region are dominantly organized in few large SCCs whose sizes follow a broad distribution and that mesostates stable at larger strains tend to be part of smaller SCCs.  

Turning now to the mesostates placed outside the reversibility region, it can be seen from Fig~\ref{Fig3}(c) that these are more likely to have one or more outgoing irreversible transitions, {\em i.e.} $k_{\rm REV} = 1, 0$. Note, that mesostates in this regime are a mixture of (i) mesostates stable at low strain values, which, however, experienced large strains in their history and subsequently were driven back to lower strains, and (ii), mesostates stable at large strains. The choice of plotting these mesostates against maximal strains in their deformation history blurs this distinction. However, we checked that the mesostates of (i) are part of some other regions of reversibility and also similarly organized into larger SCCs. On the other hand, mesostates in (ii) must belong to comparatively small SCCs, as indicated by the inset of Fig~\ref{Fig3}(d). 
In order to support our expectations regarding the mesostates of type (i), we have inspected the deformation history of mesostates belonging to an SCC, counting the number $n_Y$ of times the magnitude of the stress exceeds the yield stress (the stress at the yielding/irreversibility transition) $\sigma_y \approx 2.5$ in their deformation history. We find that this number is nearly constant across all mesostates constituting an SCC, differing only from SCC to SCC. 
Fig.~\ref{Fig_interSCC_new} shows a large excerpt of the inter-SCC graph, {\em cf.} Fig.\ref{Fig1}(c), obtained from catalog \#1. Shown is the sub-graph of $3228$ SCCs (squares) that can be reached from the SCC of the initial state by at most $15$ inter-SCC transitions. These SCCs contain a total of $12790$ mesostates. The size of the SCCs is indicated by the legend in the top left corner of the figure. The coloring scheme of the SCCs shown on the top left indicates the number $n_Y$ of yield events in the plastic deformation history of the mesostates constituting the SCC. The figure shows patches of SCCs with the same number of yield events. Among these the 'green' patch of SCCs whose constituting mesostates have suffered no yield experience is dominant. Note that even for $n_Y = 2$ (orange) or more yield events, there are relatively extended patches of SCCs of large sizes, such as the orange patch in the top left part of the graph. 

Putting all these findings together, we conclude that the landscape of local energy minima accessible by arbitrary plastic deformation protocols is composed of regions of reversibility with few but relatively large SCCs.
The mesostates belonging to these patches tend to have a common deformation history, that differs from mesostates belonging to other reversibility regions by the near-yield or yield events they suffered in their deformation history. These reversibility regions are surrounded by significantly smaller SCCs containing mesostates stable at strain values closer to yield.  

\subsection{Response to cyclic shear}

Our findings on the topology of the energy landscape, and its organization into patches of regions in which plastic events are reversible, have direct implication for the response of the amorphous solid to an applied oscillatory shear strain. An evolution towards cyclic response, {\em i.e.} limit-cycles, is a mechanism to encode memory of the past deformation history in such systems \cite{keim2019memory} and the loss of the capability to do so at increasingly larger amplitudes is believed to be related to the reversibility/irreversibility transition. We start with the observation that the mesostates forming the cyclic response to an applied oscillatory shear are mutually reachable and therefore must belong to the same SCC: consider a simple cycle with a  lower endpoint $R$, {\em i.e.} a mesostate $R$, such that 
\begin{equation}
R = \Dn^{m}\Up^{n}R\,.
\label{eqn:cycleCondition}
\end{equation}
The intermediate states of this cycle are the mesostates $R, \Up R, \Up^2 R, \ldots \Up^n R, \Dn \Up^n R, \Dn^2 \Up^n R, \ldots, \Dn^m \Up^n R = R$. Any pair of these states is mutually reachable, and these states must be part of the same SCC. Indeed, we find that many SCCs of our catalog, contain cycles of the form \eqref{eqn:cycleCondition}. 
Fig.~\ref{Fig1}(b) shows three different cycles that belong to the yellow SCC. The $\Up$- and $\Dn$- transitions forming the cycles have been highlighted by black and red arrows, respectively. 
For the state labeled $R$ we have $R = \Dn^{12}\Up^{13}R$: the amorphous solid returns to state $R$ after a sequence of 13 plastic events under increasing strain followed by 12 plastic events under decreasing strain. A cyclic response to oscillatory shear in which the period of the driving and response coincide (harmonic response) must be of the form \eqref{eqn:cycleCondition}, and will be produced by an applied cyclic shear such that  $\gamma_{\rm min} \to \gamma_{\rm max} \to \gamma_{\rm min} \to \ldots$, for some pair of  strains $\gamma_{\rm min}$ and $\gamma_{\rm max}$. 
To relate the length of a limit cycle $\ell = m + n$, {\em cf.} \eqref{eqn:cycleCondition}, to the drive amplitude, we extract from our catalog every possible limit-cycle of the form \eqref{eqn:cycleCondition} that is compatible with oscillatory forcing given by 
\begin{equation}
 0 \to \gamma \to -\gamma \to \gamma \ldots, 
 \label{eqn:cyclicShear}
\end{equation}
for some amplitude $\gamma$. 
Across the $8$ catalogs, we identified a total of $44642$ distinct limit-cycles. 
Grouping these limit-cycles by their length $\ell$, we show in Fig.~\ref{Fig4}(a) the range of amplitudes $\gamma$ for which they were observed (horizontal red line) along with their average amplitude (blue dots).
As expected, we find that the length $\ell$ of the cycle increases with the amplitude of oscillatory shear. This behavior is well described by a power-law with an exponent of $2.5$, as indicated in the figure.

\subsection{Transient response and the reversibility/irreversibility transition}

\begin{figure}[t]
\begin{center}
\includegraphics[width=\columnwidth]{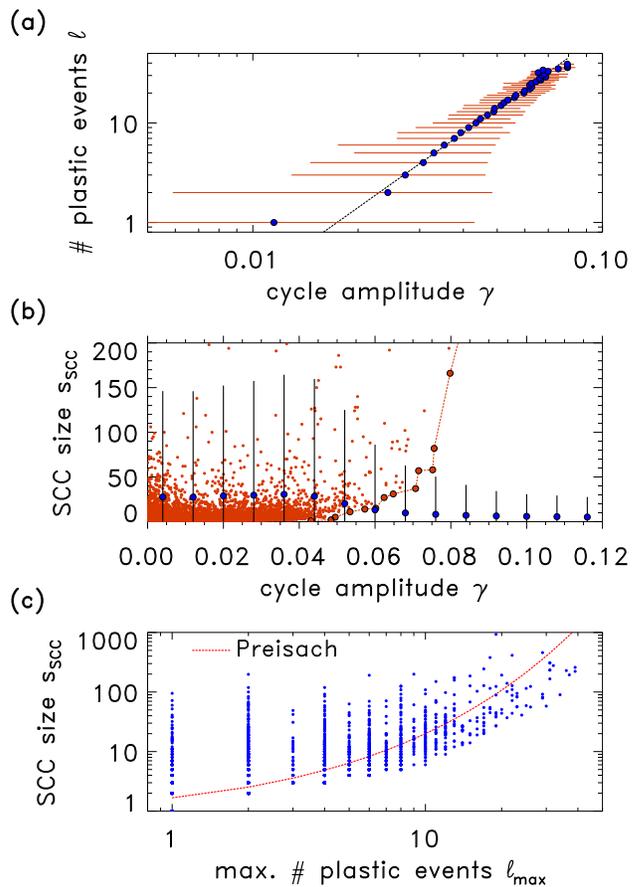}
\caption{ 
(a) The range of strain amplitudes of oscillatory shear of the form \eqref{eqn:cyclicShear} that give rise to a limit-cycle consisting of $\ell$ plastic events/mesostate transitions. Blue dots refer to average strain values for the corresponding cycle lengths. The solid 
black line is a power-law with exponent $2.5$. (b) and (c): Note that mesostates forming a cyclic response must belong to the same SCC, and consequently a limit-cycle formed by $\ell$ mesostates can only be part of an SCC whose size $s_{\rm SCC} \ge \ell$. (b) Scatter of SCC size $s_{\rm SCC}$, against the strain amplitude $ \gamma $ generating the limit-cycle with largest $\ell$ contained in the SCC (red small dots). 
The bigger red dots connected by a dashed line outline the boundary of this region, they are the smallest SCCs that contain a limit-cycle of a strain amplitude $\gamma$. Blue dots with error bars re-plot the inset of \ref{Fig3}(d), displaying the average sizes of SCCs that contain states stable at strain $\gamma$. (c) Scatter plot of the SCC sizes against the length $\ell_{\rm max}$ of the largest limit cycles, under oscillatory shear \eqref{eqn:cyclicShear} that these contain. The red curve is a prediction of the Preisach model. Refer to text for further details.  
The results in this figure combine data sampled from all $8$ catalogs.
} 
\label{Fig4}
\end{center}
\end{figure}

From the topology of the state transition graph we can draw now conclusions about the nature of the transients towards cyclic response under oscillatory shear. 
Since limit-cycles attained at increasingly larger amplitudes are formed by a larger number $\ell$ of mesostates, these require increasingly larger SCCs that can contain them since a limit cycle is always a part of an SCC. 
In Fig~\ref{Fig4}(b), we show a scatter plot where each dot corresponds to an SCC, and the size of the SCC is plotted against the strain amplitude of the limit cycle with largest $\ell$ that it contains (small red dots). We have also indicated the right boundary of the scatter region marked by a dotted red line connecting the extreme data points
\footnote{More specifically, for each possible SCC size $s$ we determined the largest amplitude $\gamma$, and from these pair of points $(s,\gamma)$ we extracted the boundary curve as 
 $s_{\rm max}(\gamma) = \max_{\gamma' \le \gamma} \{s': (s',\gamma') \}$, {\em i.e.} the convex hull.}. 
In Fig~\ref{Fig4}(b) we have superposed the data points of the inset of Fig~\ref{Fig3}(d), enabling us to compare the available SCC sizes 
with the sizes of the actual SCCs selected by the limit-cycles reached with oscillatory shear at amplitude $\gamma$. It is apparent that for amplitudes above $0.08$ the sizes of 
selected SCCs are multiple standard deviations away from the sizes of the available SCCs. This means that for these driving amplitudes, SSCs with sizes necessary to contain them are rare and thus transients are expected to be long, as observed in simulations  \cite{regev2013onset}.

%
The above observations have implications for the length of transients towards limit-cycles under oscillatory shear. They suggest that two separate dynamics govern the transient response. At low driving amplitudes, and hence well within the reversibility region, sufficiently large SCCs are abundant and cyclic response sets in when a suitable sequence of reversible plastic transitions has been reached and the SCC has thereby "trapped" the limit-cycle. Here SCC size is not a limiting factor. On the other hand,  for larger amplitudes, {\em i.e.} amplitudes outside of the reversibility region but still below yield, larger SCCs are needed, which as we have shown, become increasingly rare. It is thus plausible to assume that the ensuing increase in the duration of the  transient is predominantly due to the search for a sufficiently large SCC, and that the additional requirement that such an SCC, once found, is also trapping is of secondary importance, given that the probability of finding an SCC of the right size is already very small. These observations are consistent with earlier findings by one of us for this system which showed that limit-cycles for strain amplitudes beyond $\gamma \sim 0.13$ were not observed or extremely rare \cite{regev2013onset}. This further suggests that the reversibility/irreversibility transition of the response under oscillatory shear is governed by a cross-over of the probability of finding a limit-cycle into a rare-event regime due to the scarcity of SCCs of sufficient size. 

Having established the relation between the SCC size $s_{\rm SCC}$ and the driving amplitudes $\gamma$, we next connect $s_{\rm SCC}$ to the length of the limit cycles that they contain.
Fig~\ref{Fig4}(c) shows a scatter plot of SCC sizes against the length $\ell_{\rm max}$ of the largest limit-cycles they contain. As the figure reveals, the scatter points cover an area with a well-defined lower-boundary, {\em i.e.}  the smallest SCC size that can confine a limit-cycle of a given length $\ell$. Moreover, this boundary has a distinct concave-up shape, and for most of the data points $s_{\rm SCC} >  \ell_{\rm max}$. Thus while SCCs of size $s_{\rm SCC} =  \ell_{\rm max}$ would have sufficed to trap a limit-cycle, we find that these SCCs contain many more states in general. As will be discussed in the following section, this is also
related to the memory capacity of an SCC.

\subsection{SCCs and Memory Capacity}
To understand the origin of these excess mesostates and their connection with memory capacity, let us return to Fig.\ref{Fig1}(b) and consider the ``yellow'' SCC. This SCC is bounded by a cycle with endpoint $R$, which contains multiple sub-cycles, some of which have been indicated in the figure. It therefore appears that the largest cycles inside an SCC come with a collection of sub-cycles, the mesostates of which are mutually reachable as well. 
In fact, if the sheared amorphous solid had return point memory (RPM) \cite{sethna1993hysteresis, munganterzi2018}, then any cycle of the form \eqref{eqn:cycleCondition} would necessarily be organized in a hierarchy of sub-cycles, and moreover, all of these together would be part of the same SCC \footnote{In the context of RPM such SCCs are formed by a largest cycle and its sub-cycles. They have been referred to as maximal loops \cite{munganterzi2018}.}. 
Thus if RPM were to hold, the mesostates forming the sub-cycles of a limit-cycle would all be part of the same SCC. RPM can be used as a means to store information by utilizing the hierarchy of cycles and sub-cycles \cite{perkovic1997improved}. Moreover, the nesting depth of the hierarchy provides an upper limit for the amount of information that can be encoded via RPM \cite{perkovic1997improved,terzi2020state}.  

A central finding of ref.~\cite{mungan2019networks} has been that for amorphous solids and up to moderately large strain amplitudes, the limit-cycles reached under oscillatory shear exhibit near, but not perfect, RPM. As a result, such cycles are still accompanied by an almost hierarchical organization into sub-cycles \footnote{Fig. 2(d) of \cite{mungan2019networks} contains an example of such an organization of cycles and sub-cycles.}.
The deviations from RPM are a result of positive and negative type of interactions among soft-spots via the Eshelby deformation kernels, as a result of which a plastic event in one part of the sample may bring some soft-spots closer to instability, while at the same time it may stabilize others. If such soft-spot interactions were completely absent (or negligible), we would be in the Preisach regime, where each soft-spot can be regarded as an independent hysteretic two-state system and the system exhibits RPM \cite{Preisach1935, Barker1983}. Limit-cycles then become Preisach hysteresis loops whose cycle and sub-cycle structure is governed by the hysteresis behavior of the individual soft-spots undergoing plastic deformations as the cycle is reversed. Since the Preisach model exhibits RPM, its main hysteresis loop along with its sub-cycles constitutes an SCC. Due to the absence of interactions, the size of this SCC can be estimated as follows. Assuming that a Preisach loop is formed by $L$ non-interacting soft-spots, so that $\ell = 2L$, and assuming further that the switching sequence of soft spots as the loop is traversed is maximally random 
\footnote{By this we mean that if we label the soft-spots according to the sequence in which they undergo a plastic event as the driving force is increased, they will revert their states in some order relative to this. The latter sequence can be thought of as a permutation of the former. In \cite{terzi2020state}, we have shown that for the Preisach model this permutation alone determines the structure of the SCC containing the main hysteresis loop and hence its size. Assuming that this permutation is uniformly selected from all possible permutations, {\em i.e.} that this permutation is maximally random, the result \eqref{eqn:PreisachSCC} follows.}, 
the average size of the SCC containing the Preisach loop is given for large $L$ as \cite{terzi2020state}
\begin{equation}
    s_{\rm Pr}= \frac{1}{2} \sqrt{\frac{1}{e \pi}}\, \frac{e^{2 L^{\frac{1}{2}} }}{L^{\frac{1}{4}}}. 
    \label{eqn:PreisachSCC}
\end{equation}
In Fig~\ref{Fig4}(c), we have superposed the Preisach prediction $s_{\rm Pr}$, assuming that $L = \ell_{\rm max}/2$ in \eqref{eqn:PreisachSCC}, on top of the simulation results. Despite of the rather crude assumptions made by identifying the SCCs of the sheared amorphous solid as Preisach loops, the Preisach prediction broadly follows the lower boundary of the scattered points, {\em i.e.} the minimum size of SCCs that can support a limit-cycle of a given length $\ell_{\rm max}$. 

As remarked above, the capacity for encoding memory using RPM is related to the nesting depth $d$ of the hierarchy of cycles and sub-cycles. For the Preisach model, the average of $d$ can be worked out explicitly and is given as $d = 2 \sqrt{L}$ \cite{terzi2020state}. Comparing with the corresponding average SCC size $s_{\rm Pr}$ of \eqref{eqn:PreisachSCC}, this gives $d = \ln s_{\rm Pr}$ to leading order.  

\section{Discussion}
We have analyzed the structure of transition graphs characterizing the plastic response of an amorphous solid to an applied external shear. We have focused on the strongly connected components (SCCs) of these graphs. Physically, SCCs correspond to collections of stable particle configurations that are interconnected by reversible plastic events, so that it is possible to reach any of these configurations from any other one by an appropriate sequence of applied shear strain. The identification of SCCs thereby enabled us to designate plastic events as reversible and irreversible, depending on whether these connect states within the same SCC or not. 
The description in terms of SCCs has also allowed us to characterize the topology of the underlying energy landscape. Our analysis shows that the energy landscape is highly heterogeneous: basins of few but large SCCs, containing a large number of reversible transitions at strain values below yield, the reversibility regions, are surrounded by a large number of very small SCCs, consisting of local minima stable at strain values near or above yield. The overall size distribution of SCCs is therefore rather broad, and we find it to follow a power-law. Since the plastic events constituting any cyclic response to an applied shear must be confined to a single SCC, and the number of such plastic events increases with the amplitude of the driving, the size of the corresponding confining SCCs becomes larger as well. We have shown that as the driving amplitude approached yielding, the sizes of the required SCCs become so large that encountering SCCs of sizes that can still confine them become rare events. This observation provides a mechanism for the irreversibility transition and the associated yield strain, above which amorphous solids under slow oscillatory shear cannot find limit-cycles and the dynamics becomes irreversible.

To summarize, the graph-theoretical analysis  of the driven dynamics of amorphous solids under athermal conditions presented here, reveals new features of the energy landscape of glasses, which are responsible for the memory properties of the system. 
Furthermore, since a transition from reversible plasticity, that allows the system to store memory of past deformations, to irreversible plasticity, which allows the system to ``forget'' past deformations, is at the heart of the yielding transition, this analysis provides a new framework for understanding this transition. 
By identifying the SCC as a basic entity that groups reversible plastic events, our study also provides the basis for predicting the memory storage and retrieval capability of such systems, a topic of interest in recent experimental work on this topic \cite{keim2020global,mukherji2019strength}. 

There are many open questions that are still to be addressed. Specifically, how network features are affected by the preparation protocol of
the initial configuration and by system size and how shearing in different orientations affects the configurations encountered. 
Recent studies \cite{yeh2020glass,bhaumik2021role} have shown that amorphous solids prepared by equilibrating supercooled liquids to very low temperatures, are ``ultra-stable'' in the sense that their response is almost purely elastic up to the point of yielding (the irreversibility transition). In such
materials, preliminary results indicate a much simpler topology in the sub-yield regime. When the system size is increased, we expect the opposite - that the graph will become more complex. One can also compose different networks that stem from the same initial configuration but are rotated by an angle as was performed in recent experiments \cite{schwen2020embedding}. One can expect that despite the different orientation there will be some overlap between the networks. However, this is still to be checked against simulation data and will be the focus of future work.

{\bf Acknowledgments}
MM and IR would like to thank Sylvain Patinet, Umut Salman, Lev Truskinovsky, and Damien Vandembroucq for stimulating discussions during their stay at ESPCI as part of the Joliot Chair visiting fellowship. IR would also like to thank Asaf Szulc for useful discussions. 
MM was funded by the Deutsche Forschungsgemeinschaft (DFG, German Research Foundation) under Projektnummer 398962893, the Deutsche Forschungsgemeinschaft (DFG, German Research Foundation) - Projektnummer 211504053 - SFB 1060, and by the Deutsche Forschungsgemeinschaft (DFG, German Research Foundation) under Germany’s Excellence Strategy - GZ 2047/1, Projekt-ID 390685813.
IR was supported by the Israel Science Foundation through Grant No. 1301/17. 
SS acknowledges support through the J. C. Bose Fellowship (JBR/2020/000015), SERB, DST, India.

{\bf Author contributions}
MM and IR contributed equally to the research of this work. IA contributed in the earlier stages of the work. KD, MM, IR and SS analyzed the results and wrote the paper.

\appendix
\section{Materials and Methods}

\subsection{Sample Preparation}
\label{supp:sample_prep}

We simulated a binary system of $1024$ point particles interacting by a soft, radially-symmetric, potential described in \cite{lerner2009locality} where half of the particles are $1.4$ larger than the other half. 
For each realization, we prepared an initial configuration at a high temperature of $T=1.0$ and ran it for $20$ simulation time units (all units are mentioned in standard Lennard Jones, dimensionless reduced units). We then ran the final configuration for another $50$ simulation time units at $T=0.1$. 
This quench protocol is identical to the one used in \cite{regev2013onset} and leads to a relatively soft glass (without a stress-peak). 
The final configuration was then quenched to zero temperature using the FIRE minimization algorithm. Such a configuration is part of a mesostate and we denote this mesostate by $O$ and call it the reference state. 

Next, we applied shear to the quenched configuration under athermal quasi-static (AQS) conditions, increasing the strain by small strain steps of $\delta\gamma=10^{-4}$. The straining is implemented by means of the Lees-Edwards boundary conditions \cite{ni2019yield}, and after each step we minimize the energy of the sheared configuration using the FIRE algorithm \cite{bitzek2006structural}. 
Further details of the system and simulation can be found in \cite{lerner2009locality, regev2013onset}.

\subsection{Extraction of mesostate catalogs}
\label{supp:catalog_extraction}

Starting from the initial mesostates $O$ that we obtained from the zero temperature quench, we continue applying the strain until reaching the first plastic event. This event corresponds to a $\Up$ transition from the initial configuration $O$ at strain $\gamma^+[O]$. 
Similarly, we rerun the simulation starting from the same initial configuration and shear in the same manner but in the negative direction until another plastic event occurs, which corresponds to a $\Dn$ transition from $O$ at strain $\gamma^-[O]$. This completes the first step of obtaining $\Up O$ and $\Dn O$, forming the states of generation 1. Next, for each of the states $\Up O$ and $\Dn O$ we determine their stability ranges $\gamma^\pm[\Up O]$, $\gamma^\pm[\Dn O]$, as well as the states they transit under $\Up$ and $\Dn$, which constitute the states of generation $2$.  We then proceed in a similar manner to generation $3$ {\em etc}. After each transition we check whether the resulting mesostate has been encountered 
before or not. In the former case we just update a table of transitions, in the latter case we add the state to our collection of mesostates, which we call the {\em catalog} of mesostates. Proceeding in this way generation by generation, we assemble a catalog of mesostates $A$, their stability ranges $\gamma^\pm[A]$, along with the $\Up$ and $\Dn$ transitions among them, establishing in this way the AQS state transition graph. 

We can also associate with each mesostate $A$ the generation $g[A]$ at which it was added to our catalog.  
We quantify the extent of a catalog by the maximum number of generations $g_{\rm max}$, for which all transitions (both $\Up$- and $\Dn$-transitions) have been worked out. In other words, for all mesostates in generation $g \le g_{\rm max}$ we have identified the mesostates that they transit into. 

We have generated $8$ initial states $O$ from molecular dynamics simulations, as described above, and used these to extract  the $8$ catalogs.
Table \ref{tab:g002data} shows the number of states $N$, generations  $g_{\rm max}$ and the number of SCCs $N_{\rm SCC}$  contained in each 
catalog along with the overall totals. Along with this data, we have collected for each mesostate $A$ in our catalog, the minimum and maximum values of strain over which a mesostate is stable, as 
well as the values of the stress and energy at these two points and the changes of these two quantities when a plastic event occurs.  The analysis in the main text is based on this data  set. 
\begin{table}[h!]
\rowcolors{2}{white}{black!20}
\begin{center}
 
\begin{tabular}{ *4l }  \toprule \hline
Run   & $g_{\rm max}$ & $N$  & $N_{\rm SCC}$  \\ \midrule \hline 
1 & 40  & 48204 & 18887 \\ 
2 & 43  & 56121 & 27702 \\ 
3 & 37  & 43951 & 17451 \\ 
4 & 36  & 43550 & 18267 \\ 
5 & 41  & 44656 & 19971 \\ 
6 & 35  & 51784 & 27133 \\ 
7 & 41  & 51741 & 21516 \\ 
8 & 45  & 46395 & 18122 \\ 
 \midrule \hline 
{\bf ALL} &  n/a  & 386402 & 169049 \\ \bottomrule
\hline
\end{tabular}
\end{center}
 \caption{Properties of the $8$ mesostate catalogs, labeled 1 -- 8 that were extracted form the molecular simulations. For each catalog we list the maximum number of generations $g_{\rm max}$. This means that the catalog contains all mesostates that can be reached from the initial mesostate $O$ by at most $g_{\rm max} + 1$ plastic events, {\em i.e.}  $\Up$- or $\Dn$-transitions. In the third column we specify the number $N$ of mesostates belonging generations $g \le g_{\rm max}$, while the fourth column lists the number of strongly connected components $N_{\rm SCC}$ that these states form. }
  \label{tab:g002data}

\end{table}

The identification of the generation $g[A]$ at which a mesostate was added to our catalog also allows us via back-tracking to determine 
the deformation history, {\em i.e.} the sequence of $\Up$- and $\Dn$-transitions that lead from the initial state $O$ to $A$. A sample 
deformation history has been shown in \ref{Fig3}(b). By construction, the generation $g[A]$ is also the smallest number of 
$\Up$- and $\Dn$-transitions needed to reach $A$ from $O$. However, such a deformation history need not be unique:
with $g[A]$ being the first generation at which mesostate $A$ appears in the catalog, $A$ must necessarily have been reached with  
a transition from a mesostate belonging to generation $g-1$. However, there might be different mesostates in generation $g-1$ 
that transit into $A$, therefore each of these would provide an alternative path from $O$ to $A$. 
We have verified that such degeneracies constitute a small fraction, about 1 - 3 \%, of the transitions in our catalog.

\subsection{Identification of strongly connected components, reversible and irreversible transitions}
\label{supp:scc_identification}

Once the catalogs of mesostates have been compiled, we used an implementation of the Kosaraju-Sharir algorithm \cite{aho1982data} to identify the strongly connected components (SCC) of the transition graphs. Thus given a catalog, we are able to assign each of its mesostates to an SCC and thereby obtain their sizes. As discussed in the main text, transition between mesostates belonging to the same SCC are reversible, while those between mesostates belonging to different SCCs are irreversible. 

From a numerical implementation point of view, in which we only sample a finite number of mesostates and transitions, it is possible that a transition that appears irreversible, turns reversible in a larger catalog of mesostates, as more mesostates and transitions are added. Such conversions do indeed occur, but we find that they happen predominantly at low generations. As yielding is approached, a large number of small SCCs are generated, and the transitions between these typically involve large plastic events. It is therefore unlikely that some of these transitions will become reversible, and we verified this, inspecting our data. This is consistent with the results of Fig.~\ref{Fig3}(a) that the SCC size distribution changes little as catalogs with an increasingly larger number of generations are sampled.

Note that if a mesostate has two outgoing irreversible transitions, it must necessarily constitute an SCC of size one, {\em i.e.} an SCC containing just this mesostate. SCCs of size one also arise when a node in our catalog is {\em peripheral}, {\em i.e.} it belongs to generation $g_{\rm max} + 1$, and both outgoing transitions are left undetermined and hence absent. Since this is an artifact of the catalog acquisition procedure, we have excluded all peripheral nodes in our analysis of SCCs. 

The inter-SCC graph shown in Fig.~\ref{Fig1}(c) was obtained by collapsing the SCCs into a single vertex and keeping the irreversible transitions connecting mesostates belonging to the different SCCs. It is easy to see that each SCC will have at least two outgoing inter-SCC transitions. Since the AQS transitions graphs formed by only considering the $\Up$-transitions (or $\Dn$-transitions) are acyclic, and in particular collections of directed trees \cite{munganterzi2018, munganwitten2018, mungan2019networks}, inside each SCC there must be at least one $\Up$- and one $\Dn$-tree. The corresponding transitions out of their roots must necessarily be irreversible and hence lead out of the SCC.

The inset of Fig.~\ref{Fig3}(d) shows the mean SCC size that a mesostate belong to, given that it is stable at a strain magnitude 
$\vert \gamma \vert$. To this end we considered $25$ equally-spaced strain magnitudes with $ 0.000 \le \vert \gamma \vert \le 0.200$, so that the spacing between successive strain magnitudes is $\Delta = 0.008$. Given a strain of magnitude $ \vert \gamma \vert$, we consider all mesostates $A$ that are stable at $\pm \vert \gamma \vert$, so that $\gamma^-[A] < \vert \gamma \vert  < \gamma^+[A]$, or $\gamma^-[A] < -\vert \gamma \vert  < \gamma^+[A]$ holds. We next perform an average over the sizes of the SCCs that these mesostates belong to. The average SCC size and its standard deviation obtained in this way are then plotted  
against $\vert \gamma \vert$, leading to the inset of Fig.~\ref{Fig3}(d).

Note that the partition of the range of $\vert \gamma \vert$ values into equally spaced strain values with spacing $\Delta$, will cause mesostates $A$ whose stability range $\gamma^+[A] - \gamma^-[A]$ is larger than $\Delta$, to be associated with multiple and adjacent values of $\vert \gamma \vert$. The effect of this is a smoothing of the SCC size averages. We have checked that this does not effect our results significantly.
 
\bibliography{AmorphNets}

\end{document}